\documentclass[twocolumn,nopacs,floatfix,amsmath,nofootinbib,amssymb,preprintnumbers,floatfix]{revtex4}
\usepackage{longtable,lscape,booktabs}
\usepackage{txfonts}
\usepackage{overpic}
\usepackage{amssymb}
\usepackage{indentfirst}
\usepackage{feynmf}
\usepackage{slashed}
\usepackage{cases}
\usepackage{color}
\usepackage{multirow}
\usepackage{ulem}
\usepackage{subfigure}
\usepackage{graphicx,color,dcolumn,bm}
\usepackage[colorlinks,
            citecolor=blue,
            anchorcolor=red,
            menucolor=red,
            linkcolor=red,
            filecolor=red,
            runcolor=red,
            urlcolor=blue,
            frenchlinks=red]{hyperref}

\begin{document}

\title{New Types of Hydrogenlike Matter Composed of Electron(s) and Meson(s)}

\author{Jun-Feng Wang$^{a,b}$}\email[]{wjunfeng2023@lzu.edu.cn}
\author{Zhi-Feng Sun$^{a,b,c,d}$}\email[]{sunzf@lzu.edu.cn}
\affiliation {\it$^a$School of Physical Science and Technology, Lanzhou University, Lanzhou 730000, China\\
$^b$Research Center for Hadron and CSR Physics, Lanzhou University and Institute of Modern Physics of CAS, Lanzhou 730000, China\\
$^c$Lanzhou Center for Theoretical Physics, Key Laboratory of Theoretical Physics of Gansu Province, and Key Laboratory of Quantum Theory and Applications of the Ministry of Education, Lanzhou University, Lanzhou, 730000, China\\
$^d$Frontiers Science Center for Rare Isotopes, Lanzhou University, Lanzhou, Gansu 730000, China
}

\author{Zi-Yue Cui$^{A,B,C}$}\email[]{czy20010628@163.com}
\author{Cheng-Qun Pang$^{B,C,D}$}\email[]{xuehua45@163.com}
\affiliation
{
$^A$College of Physics and Electronic Information Engineering, Qinghai Normal University, Xining 810000, China
\\$^B$School of Physics and Optoelectronic Engineering, Ludong University, Yantai 264000, China
\\$^C$Joint Research Center for Physics, Lanzhou University and Qinghai Normal University,
Xining 810000, China
\\
$^D$Lanzhou Center for Theoretical Physics, Key Laboratory of Theoretical Physics of Gansu Province, and Key Laboratory of Quantum Theory and Applications of the Ministry of Education, Lanzhou University, Lanzhou, 730000, China
}
	
\date{\today}
	
\begin{abstract}
In the present work, we predict the existence of new types of hydrogenlike matter, including hydrogenlike atoms ($\pi^+e^-$, $K^+e^-$, $D^+e^-$), hydrogenlike molecular ions ($\pi^+\pi^+e^-$, $K^+K^+e^-$, $D^+D^+e^-$) and hydrogenlike molecules ($\pi^+\pi^+e^-e^-$, $K^+K^+e^-e^-$, $D^+D^+e^-e^-$). By solving the Schr\"{o}dinger equation, the binding energy of hydrogenlike atoms is obtained as $E_n=-\frac{1}{2n^2}$. For hydrogenlike molecular ions and molecules, the variational method is employed to calculate the binding energies, i.e., $E_+=-0.587$ and $E_0=-1.139$ for hydrogenlike molecular ions and molecules, respectively. And the bond lengths for hydrogenlike molecular ions and molecules are also calculated, whose values are $2.003$ and $1.414$, respectively. Here all the quantities are in atomic units for convenience. In addition, the strong interaction between the two constituent mesons is considered in our calculations, where we find that its influence on the hydrogenlike molecular ions and molecules can be neglected. Comparisons of hydrogenlike molecular ion and molecule with the systems governed by the strong interaction are made, which suggests the possible existence of doubly heavy triquark, hidden heavy-flavor tetraquarks and doubly heavy tetraquarks. Hopefully, these types of matter would be observed in the future with the improvement of accuracy in the high energy physical experiments.
\end{abstract}
	
\maketitle
	
\section{Introduction}\label{I}
The exploration of microscopic matter and its inner structure continues to be an area of great interest and challenge, deepening our understanding of nature. In particle physics, it is established that the material elements contain quarks, leptons, gauge bosons and Higgs boson, which together govern the microscopic world of particles and their production and decay properties. As the development of the high-energy physics experiments, a wide variety of hadronic states have been observed. These include conventional mesons and baryons, tetraquark states ($Z_{c}(3900)$, $T_{cc}^+$, $\cdots$), and pentaquark states ($P_{c}(4312)$, $P_{c}(4380)$, $P_{c}(4440)$, $P_{c}(4457)$, $\cdots$), see Refs. \cite{Chen:2016qju,Liu:2019zoy,Olsen:2014qna,Hosaka:2016pey,Lebed:2016hpi,Esposito:2016noz,Guo:2017jvc,Ali:2017jda,Karliner:2017qhf,Albuquerque:2018jkn,Brambilla:2019esw,Guo:2019twa,Meng:2022ozq,Bicudo:2022cqi,Liu:2024uxn,Chen:2022asf,Dong:2021bvy,Dong:2017gaw,Yang:2020atz,Drenska:2010kg,Briere:2016puj,Richard:2016eis,Mezzadri:2022loq,Johnson:2024omq} for reviews. 

In addition, various types of composite particles have also been investigated, including few-lepton systems (see Ref. \cite{Ma:2025rvj} and the references therein), muonic atoms, hadronic atoms (see the reviews \cite{Friedman:2007zza,Batty:1997zp,Gal:2016boi}), and $\pi-\mu$ Coulomb bound states \cite{Coombes:1976hi,Aronson:1982bz}. Refs. \cite{abdel1998existence,rebane2012existence,bressanini1998stability,rebane2012symmetry,richard1994stability,varga1997stability,mitroy2004stability,rebane2012stability,Bressanini:1997zz,rebane2001stability,rebane2003binding,gridnev2005proof,VanHooydonk:2004su,el2007investigation,Suzuki:1998wut,richard2002stability,armour2005stability,emami2015short,emami2021review} discussed four-body-exotic systems, including the predictions of $\pi^{+}\pi^{+}e^{-}e^{-}$ and $K^{+}K^{+}e^{-}e^{-}$ bound states \cite{rebane2012existence,rebane2012symmetry}. Hydrogenlike charm-pion or charm-kaon matter has also been predicted in Ref. \cite{Luo:2023hnp}. 

Inspired by these studies, we predict in this work new types of hydrogenlike matter composed of electron(s) and meson(s), i.e., $\pi^{+}e^{-}$, $K^{+}e^{-}$, $D^{+}e^{-}$, $\pi^{+}\pi^{+}e^{-}$, $K^{+}K^{+}e^{-}$, $D^{+}D^{+}e^{-}$, $D^{+}D^{+}e^{-}e^{-}$, as well as the previously studied $\pi^{+}\pi^{+}e^{-}e^{-}$, $K^{+}K^{+}e^{-}e^{-}$ systems in Refs. \cite{rebane2012existence,rebane2012symmetry}. For these composite systems, we not only investigate their bounding energies and the wave functions, but also discuss whether or not the strong interaction impact the bounding energies. 

With ongoing progress in experimental facilities, the production of the hydrogenlike matter containing electrons and mesons may become possible. For instance, the proposed Super $\tau$-Charm Facility (STCF) \cite{Achasov:2023gey} by the Chinese particle physics community may provide a potential platform for producing and studying these composite particles. 

Besides the intrinsic value of exploring these electron-meson systems, the reason why these composite particles are of particular significance is that they are similar to the systems composed of heavy and light quarks, potentially offering insight into such configurations. For example, a tetraquark state contains two heavy quarks and two light antiquarks, while a hydrogenlike molecule comprises two electrons and two mesons, where the mesons are much heavier than the electrons. Thus, the hydrogenlike systems can be viewed as the QED analog of quark-based systems.

This paper is organized as follows. After the introduction, the theoretical framework is described in Sec. \ref{sec2}, which contains the calculations of $Me^{-}$, $MMe^{-}$ and $MMe^{-}e^{-}$ systems ($M$ = $\pi^{+}$, $K^{+}$, $D^{+}$). After that, we show the numerical results in Sec. \ref{sec3} and give a discussion in Sec. \ref{sec4} whether the strong interaction affect the systems. In Sec. \ref{sec5}, a short summary is given.

\section{FORMALISM}\label{sec2}
In this work, we will consider three new types of matter composed of meson(s) ($\pi^+/K^+/D^+$) and electron(s). These constituent masses are similar to those of the hydrogen atom, molecular ion and molecule. In order to derive the binding energy, we use the method by analogy with the hydrogen cases, which are shown in the following.

\subsection{Hydrogenlike atoms ($\pi^+e^-, K^+e^-$,  $D^+e^-$)}

We present the calculation procedure for the hydrogenlike atom. Adopting atomic units (a.u.) for this discussion, the Hamiltonian {of the hydrogenlike atom, 
\it{i.e.}, $\pi^+e^-, K^+e^-$ \rm{and} $D^+e^-$ \rm{reads}}
\begin{eqnarray}
    H&=&-\frac{1}{2}\nabla^2-\frac{1}{r}   \nonumber\\
    &=&-\frac{1}{2}\frac{1}{r^2}~\frac{\partial}{\partial r}~r^2~\frac{\partial}{\partial r}+\frac{1}{2}~\frac{1}{r^2}~\hat{\vec{L}}^2-\frac{1}{r},
\end{eqnarray}
{where $\hat{\vec{L}}$ is the angular momentum operator, \it{i.e.},}
\begin{eqnarray}
    \hat{\vec{L}}&=&\hat{\vec{r}}\times\hat{\vec{p}}=-i\vec{r}\times\nabla \nonumber\\
    &=&-i~\bigg(\vec{e_{\varphi}}~\frac{\partial}{\partial\theta}-\vec{e_{\theta}}~\frac{1}{\sin{\theta}}~\frac{\partial}{\partial\varphi}\bigg).
\end{eqnarray}
{and}
\begin{eqnarray}
    \hat{\vec{L}}^2&=&-\bigg(\frac{1}{\sin{\theta}}~\frac{\partial}{\partial\theta}~\sin{\theta}~\frac{\partial}{\partial\theta}+\frac{1}{\sin^2{\theta}}~\frac{\partial^2}{\partial\varphi^2}\bigg).
\end{eqnarray}
Note here that in a.u., $m_e=e=\hbar=1$. After solving the stationary-state Schr\"{o}dinger equation, we drive 
\begin{eqnarray}
    E_n&=&-\frac{1}{2n^2},\\
    \psi_{nlm}(r,\theta,\varphi)&=&R_{nl}(r)~Y_{lm}(\theta,\varphi),
\end{eqnarray}
where the radial wave function is
\begin{eqnarray}
    R_{nl}(r)&=&\frac{2}{n^2(2l+1)!}\bigg[\frac{(n+l)!}{(n-l-1)!}\bigg]^{1/2}\xi^l \nonumber\\
    &&\times F(-n_r,2l+2,\xi)e^{-\xi/2},
\end{eqnarray}
the Spherical Harmonics is $Y_{lm}(\theta,\varphi)$, and 
\begin{eqnarray}
    F(a,b,\chi)&=&1+\frac{a}{b}\chi+\frac{a(a+1)}{b(b+1)}\frac{\chi^2}{2!}+\frac{a(a+1)(a+2)}{b(b+1)(b+2)}\frac{\xi^3}{3!}\nonumber\\
    &&+\cdots
\end{eqnarray}
is the Confluent hypergeometric function, $n=n_r+l+1=1,2,3,\cdots$ is the principal quantum number, $\xi=\frac{2r}{n}$, $l=0,1,2,\cdots$ is the angular momentum quantum number, $m=0,\pm1,\pm2,\cdots$ is the magnetic quantum number. Specially, the energy and wave function of the ground eigenstate are 
\begin{eqnarray}
     E_1&=&-\frac{1}{2},\\
    \psi_{100}(r,\theta,\varphi)&=&\bigg(\frac{1}{\pi}\bigg)^{1/2}e^{-r}. \label{a}
\end{eqnarray}

\subsection{Hydrogenlike molecular ions ($\pi^+\pi^+e^-$, $K^+K^+e^-$, $D^+D^+e^-$)}

According to the Born-Oppenheimer approximation, the relative distance between the two mesons can be viewed as parameter, not the dynamical variable, such that the Hamiltonian of the systems can be expressed as
\begin{eqnarray}
    H&=&-\frac{1}{2}\nabla^2-\frac{1}{r_a}-\frac{1}{r_b}+\frac{1}{R}\nonumber\\
    &=&H_e+\frac{1}{R},
\end{eqnarray}
where the spin of the electron is neglected, $r_a$ and $r_b$ are the relative distances between the electron and the two mesons, respectively,
$H_e\equiv-\frac{1}{2}\nabla^2-\frac{1}{r_a}-\frac{1}{r_b}$ depicts the motion of the electron in the two-center potential.

Like the case of hydrogen molecular ion, we adopt the variational method to calculate the binding energy of the systems. In this work, we only focus on the ground eigenstate case. In light of that the electron is only affected by the attractive Coulomb force of the two identical mesons, it is naturally to choose the trial wave function as the superposition as follows
\begin{eqnarray}
    \psi=C_{a}\psi_a+C_{b}\psi_b, \label{eq4}
\end{eqnarray}
where
\begin{eqnarray}
    \psi_a=\bigg(\frac{\lambda^3}{\pi}\bigg)^{1/2}e^{-\lambda r_a},\\
    \psi_b=\bigg(\frac{\lambda^3}{\pi}\bigg)^{1/2}e^{-\lambda r_b},
\end{eqnarray}
with $\lambda$ is the variational parameter. If $\lambda$ is fixed as $\lambda=1$, $\psi_a$ and $\psi_b$ become the ground-state wave function of hydrogenlike atom, i.e., Eq. \eqref{a}.

Since the potential of the system is invariant when performing the reflection with respect to the midpoint of the two mesons, the state of the electron can be classified according to the reflection symmetry. Here we have $C_a=C_b\equiv C_+$ for even parity, and $C_a=-C_b\equiv C_-$ for odd parity. So the trial wave function can be expressed as
 \begin{eqnarray}
    \psi_{\pm}=C_{\pm}(\psi_a\pm\psi_b). \label{eq4}
\end{eqnarray}
From the normalization condition, we obtain
\begin{eqnarray}
    C_{\pm}=\frac{1}{\sqrt{2}}\bigg\{1\pm\big[\frac{1}{3}(\lambda R)^2+\lambda R+1 \big] e^{-\lambda R}\bigg\}^{-1/2}.
\end{eqnarray}
Using Eq.~\eqref{eq4}, we get the expression for the average value of the energy
\begin{eqnarray}
    E_{\pm}&=&\langle\psi_{\pm}\mid H\mid \psi_{\pm}\rangle \nonumber\\
    &=&\langle\psi_{\pm}\mid H_e\mid \psi_{\pm}\rangle+\frac{1}{R} \nonumber\\
   &=&2C^2_{\pm} \bigg\{\frac{\lambda^2}{2}-\lambda-\frac{1}{R}+\frac{1}{R}(\lambda R+1)e^{-2\lambda R}    \\
   && \pm \big[\frac{\lambda^3 R}{2} +\frac{\lambda^2}{2}-2\lambda^2 R-2\lambda-\frac{\lambda^4 R^2}{6}\big]e^{-\lambda R}\bigg\}+\frac{1}{R}. \nonumber 
\end{eqnarray}
The variational parameters $\lambda$ and $R$ are determined by the variational extremum condition, i.e.,
\begin{eqnarray}
  \frac{\partial E_{\pm}}{\partial R}=0,~~
      \frac{\partial E_{\pm}}{\partial \lambda}=0.
\end{eqnarray}

\subsection{Hydrogenlike molecules ($\pi^+\pi^+e^-e^-$, $K^+K^+e^-e^-$, $D^+D^+e^-e^-$)}

In 1927, Heitler and London successfully applied quantum mechanics to study the structure of hydrogen molecule \cite{Heitler1927}, and pioneered the field of quantum chemistry.

Similar to the hydrogen molecule system, the hydrogenlike molecule system consists of two identical mesons and two electrons. According to Born-Oppenheimer approximation, the motion of the two mesons can be neglected, such that the hydrogenlike molecule is simplified as two-electron problem, of which the Hamiltonian reads
\begin{eqnarray}
    H&=&-\frac{1}{2}(\nabla_1^2+\nabla_2^2)-\frac{1}{r_{a1}}-\frac{1}{r_{a2}}-\frac{1}{r_{b1}}-\frac{1}{r_{b2}}    \nonumber\\
   && +\frac{1}{r_{12}}+\frac{1}{R}.
\end{eqnarray}
The distance $R$ between the two mesons is treated as a parameter in our calculation. And next we utilize the variational method to get the eigenenergy and wave function of the ground state. In view that when $R$ becomes very large, the ground-state wave function of the hydrogenlike molecule can approximately written as the product of the ground-state wave functions of the two hydrogenlike atoms, considering the exchange symmetry of the two electrons, the trial wave function can be chosen as
\begin{eqnarray}
    \psi_+(1,2)&=&\phi_0(1,2)~\chi_0(S_{1z},S_{2z}),\\
    \psi_-(1,2)&=&\phi_1(1,2)~\chi_1(S_{1z},S_{2z}),\\
\phi_0(1,2)&=&C_S\bigg[\psi(r_{1a})\psi(r_{2b})+\psi(r_{2a})\psi(r_{1b})\bigg],\\
\phi_1(1,2)&=&C_A\bigg[\psi(r_{1a})\psi(r_{2b})-\psi(r_{2a})\psi(r_{1b})\bigg],
\end{eqnarray}
where $\chi_0(S_{1z},S_{2z})$ and $\chi_1(S_{1z},S_{2z})$ are the spin wave functions of singlet and triplet, respectively. Since the Hamiltonian $H$ does not depend on spin, we do not consider spin wave function in the calculation of variational
method. Taking into account the influence on the electron of the mesons and another electron, the single-electron wave functions can be chosen as
\begin{eqnarray}
    \psi(r)=\bigg(\frac{\lambda^3}{\pi}\bigg)^{1/2}e^{-\lambda r},
\end{eqnarray}
where $\lambda$ is the variational parameter. $C_S$ and $C_A$ are determined by the normalization conditions, i.e.,
\begin{eqnarray}
     C_S&=&\frac{1}{\sqrt{2}}\bigg[1+\bigg(\frac{1}{3}\lambda^2R^2+\lambda R+1\bigg)^2e^{-2\lambda R}\bigg]^{-1/2},\\
     C_A&=&\frac{1}{\sqrt{2}}\bigg[1-\bigg(\frac{1}{3}\lambda^2R^2+\lambda R+1\bigg)^2e^{-2\lambda R}\bigg]^{-1/2}.
\end{eqnarray}
The average value of the energy are obtained as
\begin{eqnarray}
    E_{0}&=&\langle\phi_0(1,2)\mid H\mid \phi_0(1,2)\rangle  \nonumber\\
    &=&4C_s^2 I_{H1}+4C_s^2 I_{e1}+4C_s^2I_{H2}S_1+4C_s^2I_{e2}S_1  \nonumber\\
   && +2C_s^2I_{DCC}+2C_s^2I_{DCE}+\frac{1}{R},\label{eq26}\\
   E_{1}&=&\langle\phi_1(1,2)\mid H\mid \phi_1(1,2)\rangle  \nonumber\\
   &=&4C_A^2 I_{H1}+4C_A^2 I_{e1}-4C_A^2I_{H2}S_1-4C_A^2I_{e2}S_1  \nonumber\\
   && +2C_A^2I_{DCC}-2C_A^2I_{DCE}+\frac{1}{R}, 
\end{eqnarray}
where
\begin{eqnarray}
    I_{DCE}&=&\frac{6}{5R}\bigg\{-S_2^2E_i(4\lambda R)+2S_1S_2E_i(2\lambda R)+S_1^2(\gamma+\ln{(\lambda R)}) \bigg\}          \nonumber\\
  &&  +\frac{e^{-2\lambda R}}{R}\bigg[-\frac{1}{15}(\lambda R)^4-\frac{3}{5}(\lambda R)^3-\frac{23}{20}(\lambda R)^2+\frac{5}{8}\lambda R\bigg],\\
    I_{DCC}&=&-\lambda\bigg(\frac{\lambda^2R^2}{6}+\frac{3}{4}\lambda R+\frac{11}{8}+\frac{1}{\lambda R}\bigg)e^{-2\lambda R}+\frac{1}{R},    \\
    I_{H1}&=&\frac{\lambda^2}{2}-\lambda,\\
    I_{H2}&=&\bigg[\frac{\lambda^3R}{2}+\frac{\lambda^2}{2}-\lambda^2R-
    \lambda-\frac{\lambda^4R^2}{6}\bigg]e^{-\lambda R},\\
    I_{e1}&=&-\frac{1}{R}\bigg[-(\lambda R+1)e^{-2\lambda R}+1\bigg],\\
    I_{e2}&=&- \lambda(\lambda R+1)e^{-\lambda R},\\
    S_1&=&e^{-\lambda R}\bigg(\frac{1}{3}\lambda^2R^2+\lambda R+1\bigg),\\
    S_2&=&e^{\lambda R}\bigg(\frac{1}{3}\lambda^2R^2-\lambda R+1\bigg),\\
    E_i(\rho)&\equiv&\int^{\infty}_{\rho}\frac{1}{t}e^{-t}dt.
\end{eqnarray}
The parameters $\lambda$ and $R$ are also determined by the variational extremum condition,
\begin{eqnarray}
  \frac{\partial E_{0,1}}{\partial R}=0,~~
      \frac{\partial E_{0,1}}{\partial \lambda}=0.
\end{eqnarray}

\section{NUMERICAL RESULTS}\label{sec3}
\begin{figure}
\centering
\vspace{0.5cm}
\setlength{\abovecaptionskip}{0cm} 
\begin{minipage}{7cm}
\includegraphics[width=1.00\linewidth]{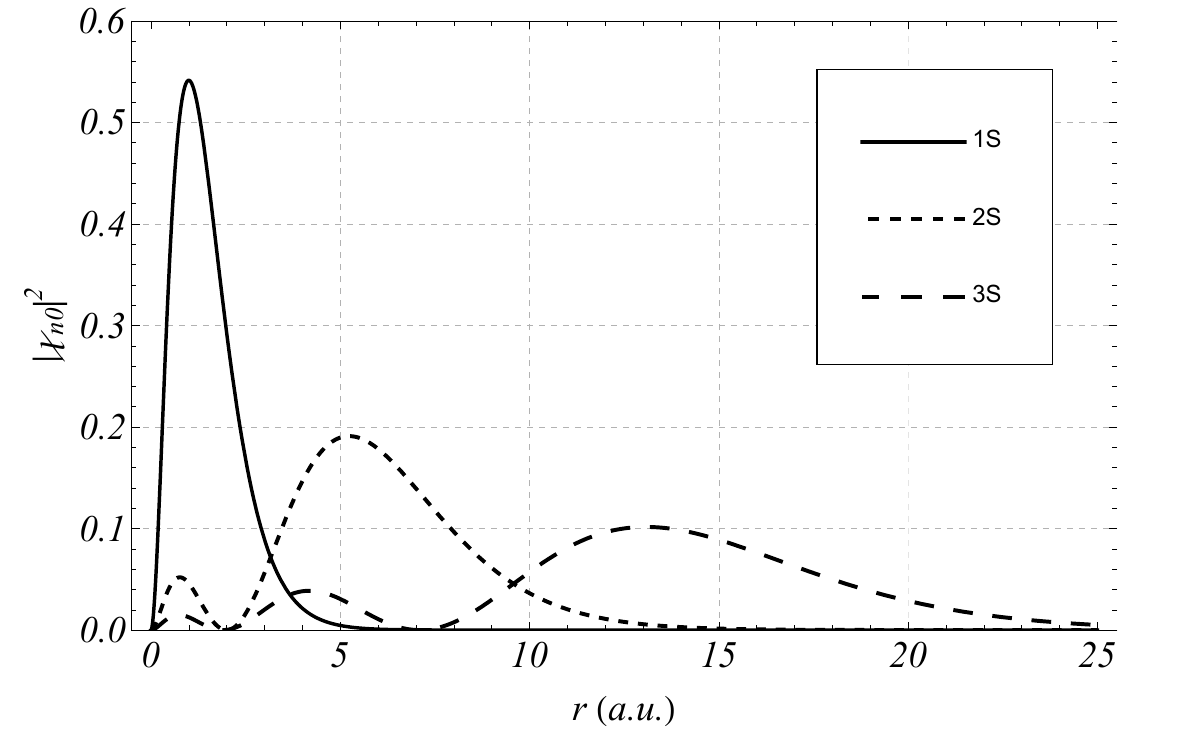}
\end{minipage}
\caption{The Radial probability density of the 1S, 2S and 3S state with $\chi_{n0}=rR_{n0}(r)$.}
\label{Fig1}
\end{figure}

\begin{figure}[htbp]
\begin{minipage}{0.49\linewidth}
\centering
\includegraphics[width=0.95\linewidth,trim=0 0 0 0,clip]{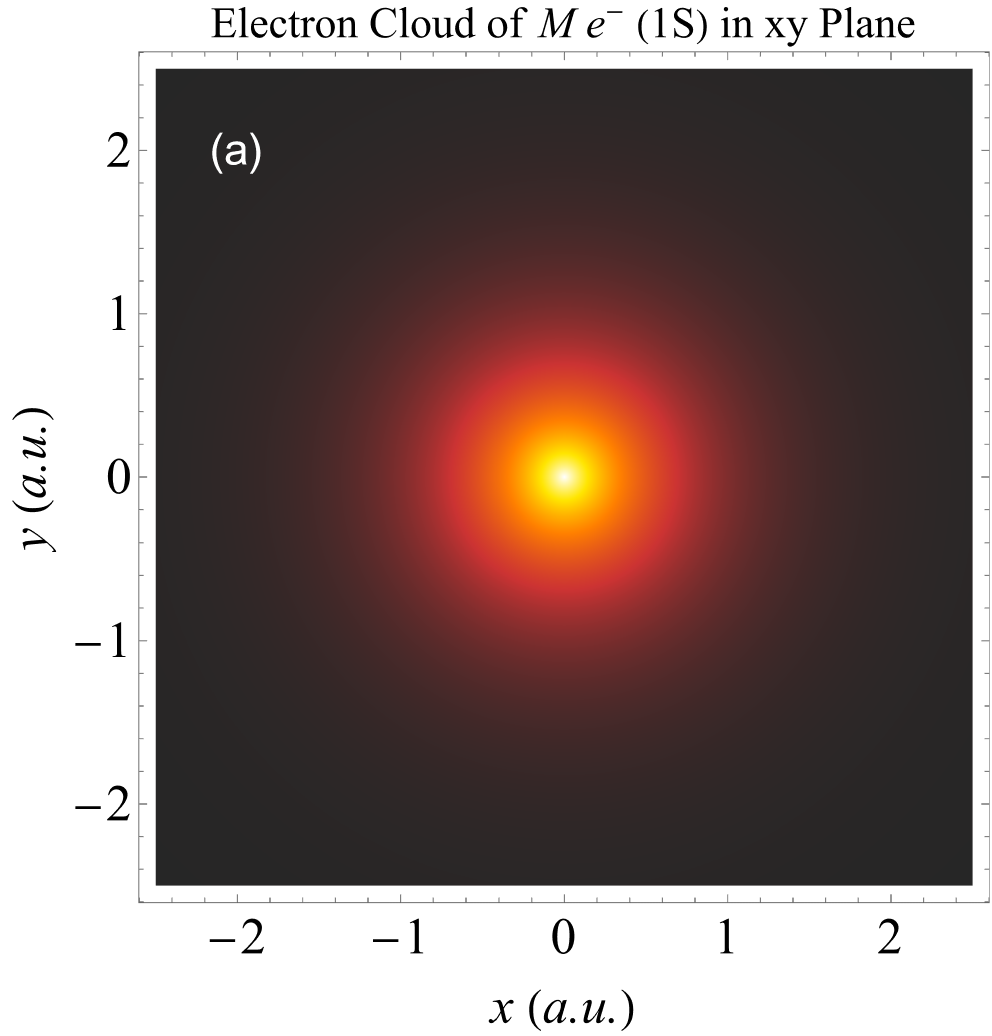} 
\label{Fig2a}
\end{minipage}
\begin{minipage}{0.49\linewidth} 
\centering 
\includegraphics[width=0.98\linewidth,trim=0 0 0 0,clip]{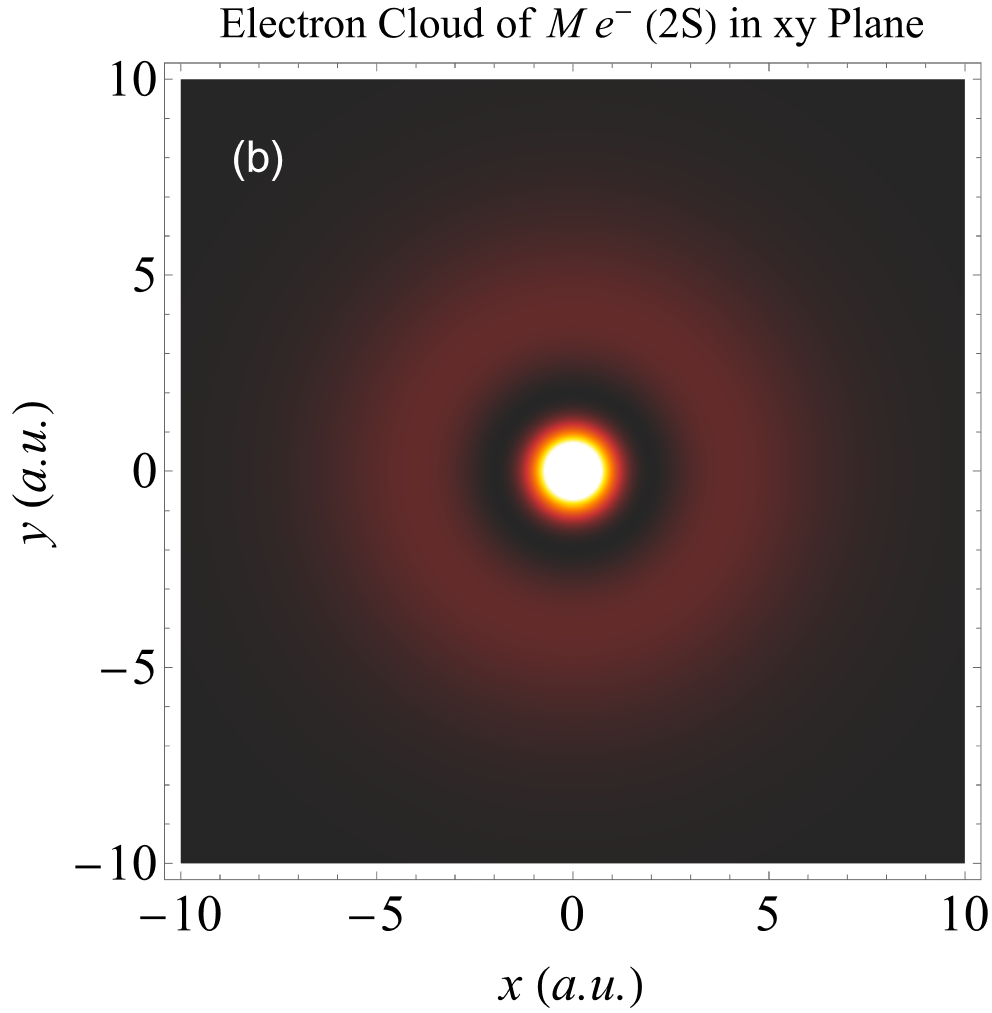} 
\label{Fig2b}  
\end{minipage}	

\vspace{3mm}

\begin{minipage}{0.49\linewidth} 
\centering 
\includegraphics[width=0.98\linewidth,trim=0 0 0 0,clip]{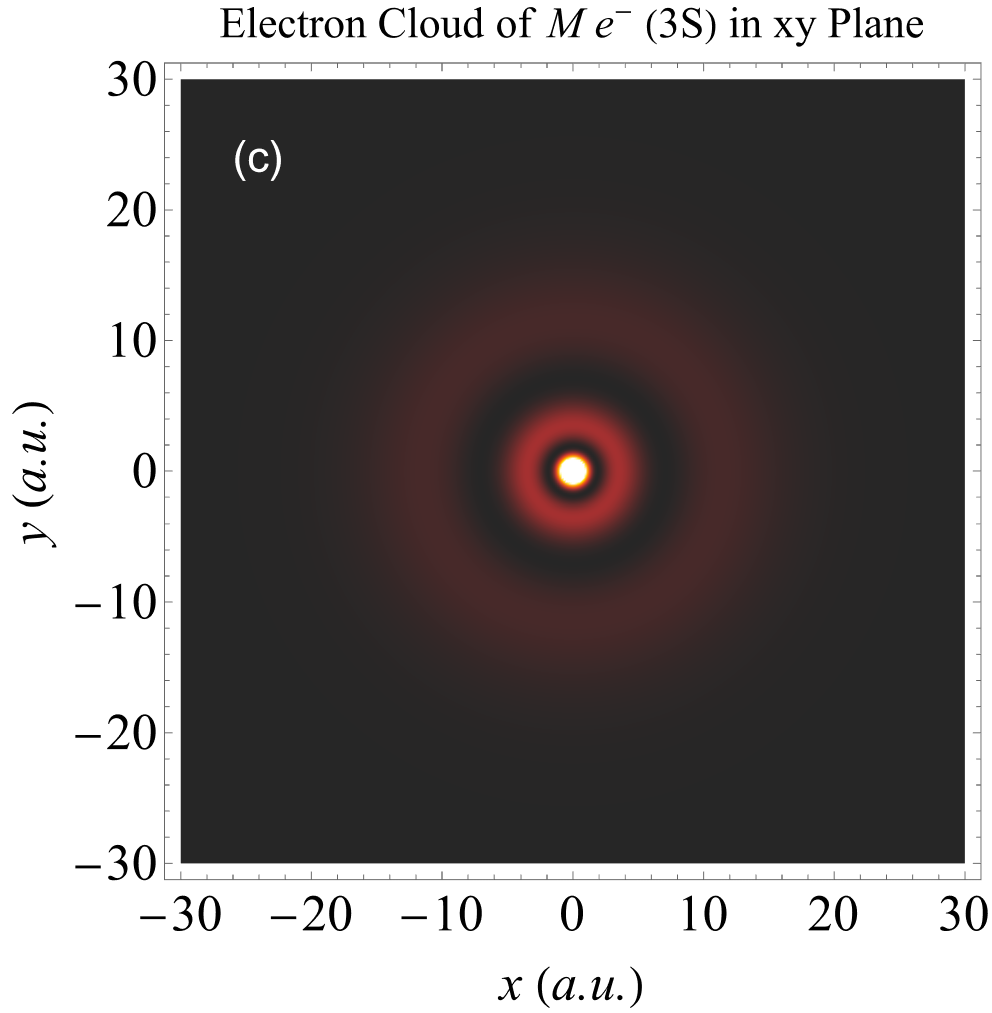} 
\label{Fig2c}  
\end{minipage}
\hspace{0.05mm}
\begin{minipage}{0.49\linewidth} 
\centering 
\includegraphics[width=0.95\linewidth,trim=0 0 0 0,clip]{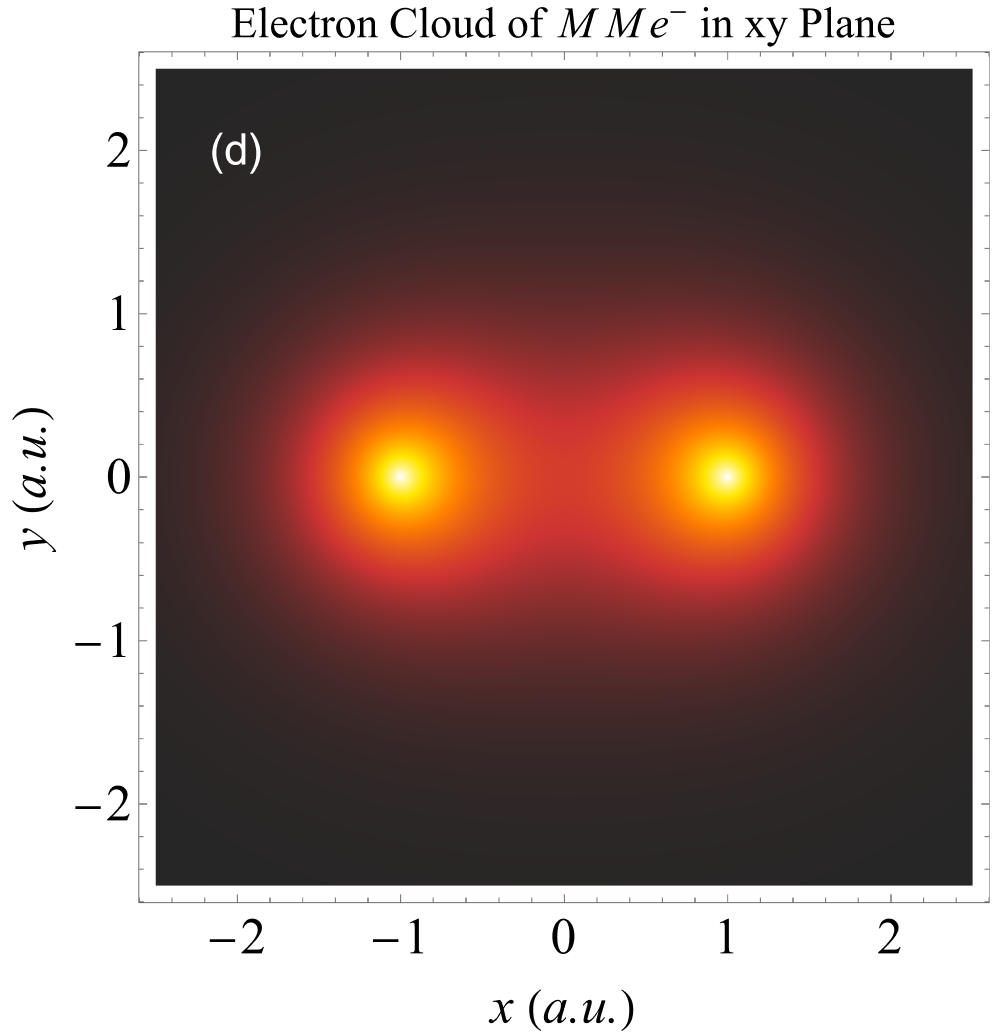} 
\label{Fig2d}  
\end{minipage}
\caption{(color online) The electron cloud of hydrogenlike atoms and molecular ions in $xy$ plane. (a), (b), (c) and (d) corresponds to $Me^-\ (1S)$, $Me^-\ (2S)$, $Me^-\ (3S)$ and $MMe^-e^-$ with $M=\pi^+,K^+,D^+$, respectively.}
\label{Fig2}
\end{figure}

In this work, we propose three types of hydrogenlike matter including $Me^{-}$, $MMe^{-}$, $MMe^{-}e^{-}$ ($M=\pi^{+},K^{+},D^{+}$). Since the mass of electron is much smaller than that of $\pi^{+}$, $K^{+}$ or $D^{+}$, these systems have great similarity with the hydrogen atom, hydrogen molecular ion and hydrogen molecule. In our calculation, we only need to replace the proton in hydrogen systems with $\pi^{+}$, $K^{+}$ or $D^{+}$. By solving the Schr$\rm{\Ddot{o}}$dinger equation, we obtain that the binding energy of $Me^{-}$ ($M=\pi^{+},K^{+},D^{+}$) system is 
\begin{eqnarray}
    E_{1}=-\frac{1}{2}, E_{2}=-\frac{1}{8}, E_{3}=-\frac{1}{18}, \cdots
\end{eqnarray}
which is the same as that of hydrogen atom, when neglecting the motion of the meson or the proton. Beside the binding energy of $Me^{-}$ systems, we also plot the radial probability density as well as the electron cloud of $n=1,2,3$ and $l=0$ state, which is shown in FIG. \ref{Fig1} and FIG. \ref{Fig2} (a), (b), (c). Due to the electromagnetic interaction between $M$ and $e^{-}$ is the same as that of proton and $e^{-}$ in hydrogen atom, the wave function of electron in $Me^{-}$ systems is consistent with that of in hydrogen atom.

\begin{figure}
\centering
\vspace{0.5cm}
\setlength{\abovecaptionskip}{0cm} 
\begin{minipage}{7cm}
\includegraphics[width=1.00\linewidth]{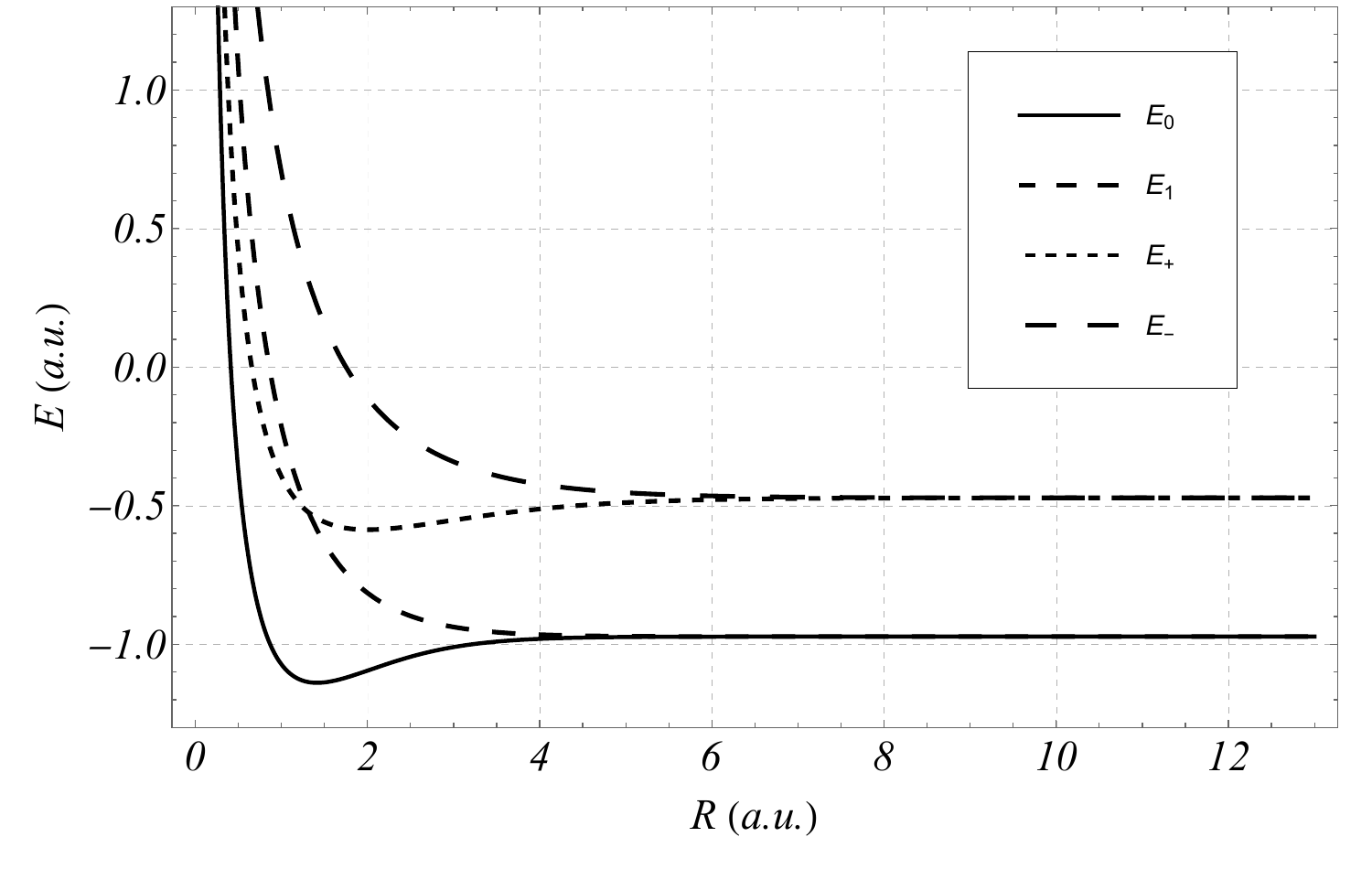}
\end{minipage}
\caption{The line shapes of the $E_{+}$, $E_{-}$, $E_{0}$ and $E_{1}$. The variational parameter $\lambda$ is fixed as 1.238 for $E_{+}$, $E_{-}$, and as 1.166 for $E_{0}$, $E_{1}$.}
\label{Fig3}
\end{figure}

For $MMe^{-}$ system of parity odd state, the average value of energy $E_{-}$ does not have minimum value. If $\lambda$ is fixed, we clearly see in FIG. \ref{Fig3} that $E_{-}$ is monotonically decreasing when the distance between the two mesons is increasing. In this case, the electron and two mesons can not form a bound state.

For $MMe^{-}$ system of even parity, there is a minimum value of $E_{+}$, i.e., $E_{+}=-0.587$, where the variational parameters $\lambda=1.238$ and $R=2.003$. So the two identical mesons and the electron can form a bound state. This can be easily understood: since $\psi_{+}$ is symmetric when exchanging the two mesons, there is a large distribution of electron near the midpoint of the mesons, forming a cloud with negative charge (see FIG. \ref{Fig2} (d)); and the electron cloud has an attractive interaction on the mesons with positive charge.

For the hydrogenlike molecule, there is no minimum value for $E_{1}$ which can be seen in FIG. \ref{Fig3}. However, $E_{0}$ has a minimum $E_{0}=-1.139$ at $\lambda=1.166$, $R=1.414$. In the limit $\lambda\to1$ and $R\to\infty$, we obtain from Eq. \eqref{eq26} that $E_{0}=-1$. The picture is that a hydrogen molecule becomes two independent hydrogen atoms in this limit, and the energy of the electrons is the sum of the binding energies ($E_{e}=-\frac{1}{2}$ for each atom) of the atoms. Despite the fact that a hydrogenlike molecule is make of two neutral hydrogenlike atom, they can form a bound state.

For hydrogenlike molecule, the spin wave function of the electron pair is antisymmetric, i.e., the two electrons are in a spin singlet, whereas the spatial wave function is symmetric. So the electrons can yet close to each other, which means they can form electron cloud with relatively higher density. And the electron cloud has a strong attractive force on the two mesons, such that the two hydrogenlike atoms combine together. The structure that the electron pair is shared by both mesons and their spins are in a singlet is so called covalent bond. 

\section{Discussion: Does the strong interaction affect the systems?}\label{sec4}

In this section, we discuss whether or not the strong interaction between the mesons impact on the systems considered in our work.

The corresponding Lagrangian are constructed by the hidden local symmetry and chiral symmetry, in which the terms needed in the present work are show below,
\begin{eqnarray}
    \mathcal{L}_{V\Phi\Phi}&=&-\frac{i}{\sqrt{2}}\langle[\Phi,\partial_{\mu}\Phi]V^{\mu}\rangle,\\
    \mathcal{L}_{VPP}&=&ia_{1}(PV_{\mu}D^{\mu}P^{\dagger}-D^{\mu}PV_{\mu}P^{\dagger}).
\end{eqnarray}
The matrices of $\Phi$, $V_{\mu}$ and $P$ read
\begin{eqnarray}
		\Phi&=&\sqrt{2}\left(
		\begin{array}{ccc}
			\frac{\sqrt{3}\pi^0+\eta+\sqrt{2}\eta'}{\sqrt{6}}&\pi^+&K^+\\
			\pi^-&\frac{-\sqrt{3}\pi^0+\eta+\sqrt{2}\eta'}{\sqrt{6}}&K^0\\
			K^-&\bar{K}^0&\frac{-2\eta+\sqrt{2}\eta'}{\sqrt{6}}
		\end{array}
		\right),
\end{eqnarray}  
\begin{eqnarray}
		V_{\mu}&=&\frac{g_{V}}{\sqrt{2}}\left(
		\begin{array}{ccc}
			\frac{1}{\sqrt{2}}(\rho^{0}+\omega)&\rho^+&K^{*+}\\
			\rho^-&-\frac{1}{\sqrt{2}}(\rho^{0}-\omega)&K^{*0}\\
			K^{*-}&\bar{K}^{*0}&\phi
		\end{array}
		\right)_\mu,\\
		P&=&\left(D^{0},D^{+},D_{s}^{+}\right). 
\end{eqnarray}
The constant $g_{V}$ is determined from the experimental value of $g_{\rho\pi\pi}$, i.e., $g_{V}=5.80\pm0.9$ \cite{Harada:2003jx}. Comparing $\mathcal{L}_{VPP}$ with the Lagrangian in Ref. \cite{Isola:2003fh}, we have 
\begin{eqnarray}
a_{1}=-\frac{\beta}{m_{P}},
\end{eqnarray}
where the parameter $\beta$ is determined by the vector meson dominance \cite{Isola:2003fh,Colangelo:1993zq}, i.e., $\beta=0.9$. $m_{P}$ is the mass of pseudoscalar charmed meson. The symbol $\langle\cdots\rangle$ is trace of a matrix. 

\begin{figure}
\centering
\vspace{0.5cm}
\setlength{\abovecaptionskip}{0cm} 
\begin{minipage}{7cm}
\includegraphics[width=1.00\linewidth]{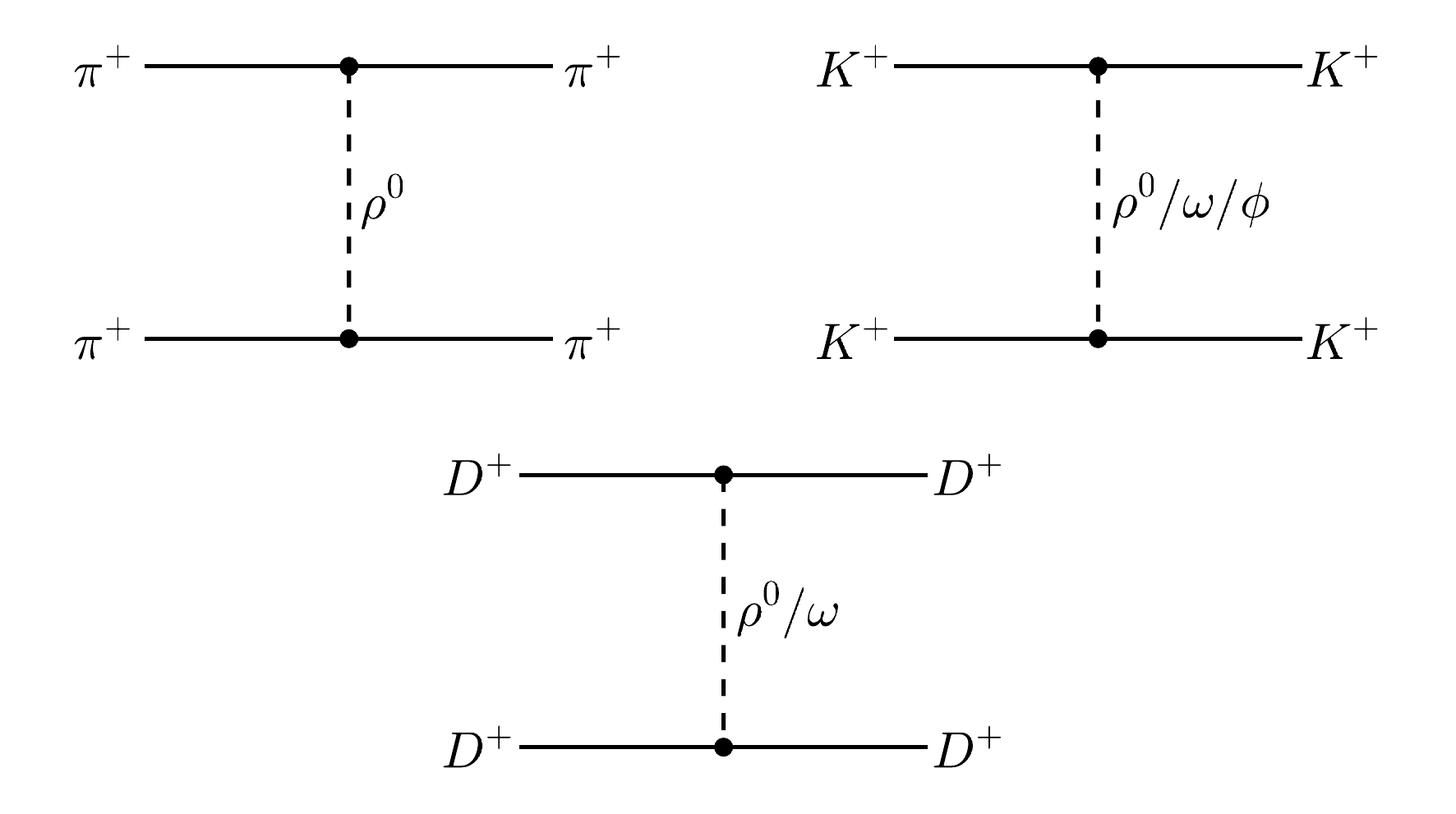}
\end{minipage}
\caption{The Feynman diagrams of the $\pi^{+}$-$\pi^{+}$, $K^{+}$-$K^{+}$ and $D^{+}$-$D^{+}$ interactions.}
\label{Fig5}
\end{figure}

\begin{figure}
\centering
\vspace{0.5cm}
\setlength{\abovecaptionskip}{0cm} 
\begin{minipage}{7cm}
\includegraphics[width=1.00\linewidth]{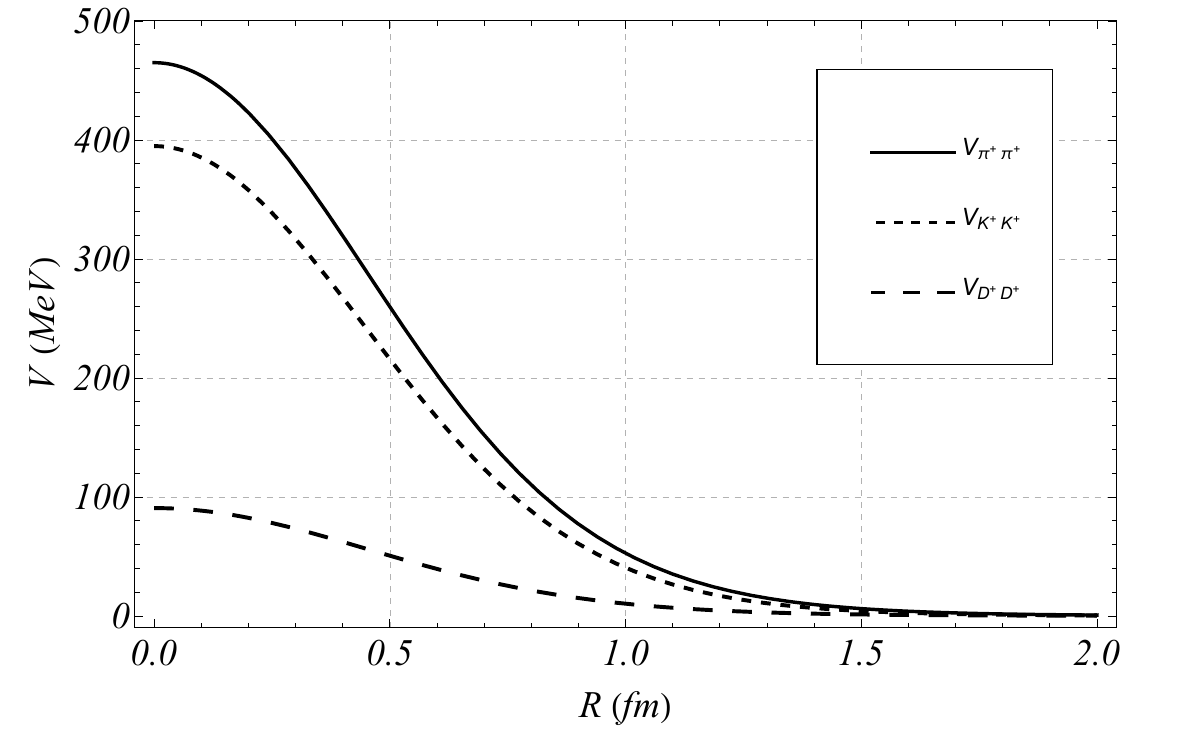}
\end{minipage}
\caption{The strong interaction potential of $V_{\pi^{+}\pi^{+}}$, $V_{K^{+}K^{+}}$ and $V_{D^{+}D^{+}}$.}
\label{Fig6}
\end{figure}

In FIG. \ref{Fig5}, we show the diagrams corresponding to the strong interactions of $\pi^{+}\pi^{+}$, $K^{+}K^{+}$ and $D^{+}D^{+}$, which occur through the exchange of $\rho^{0}$, $\omega$ and $\phi$. The effective potentials read
\begin{eqnarray}
    V_{\pi^{+}\pi^{+}}&=&2g_{V}^{2}Y(\Lambda,m_{\rho^{0}},r),\label{eq45}\\ \nonumber
    V_{K^{+}K^{+}}&=&\frac{g_{V}^{2}}{2}\left[Y(\Lambda,m_{\rho^{0}},r)+Y(\Lambda,m_{\omega},r)\right.\\
    &&\left.+2Y(\Lambda,m_{\phi},r)\right],\label{eq46}\\
    V_{D^{+}D^{+}}&=&\frac{a_{1}^{2}g_{V}^{2}m_{D^{+}}^{2}}{4}\left[Y(\Lambda,m_{\rho^{0}},r)+Y(\Lambda,m_{\omega},r)\right],\label{eq47}
\end{eqnarray}
of which the procedure of calculation is show in the appendix. Note that we have introduce the exponential form factor $ F(q^{2})=e^{(q_{0}^{2}-\Vec{q}^{2})/\Lambda^{2}}$ for each vertex in the diagrams. $\Lambda$ is the cutoff which is taken as 1956.95 (1000 $\rm{MeV}$ in the natural units) in the present work. The line shapes of the effective potentials in Eq. \eqref{eq45}-\eqref{eq47}  are shown in FIG. \ref{Fig6}.

The next step is using the varitional method, where we only need to replace the potential of the two mesons for hydrogenlike molecular ion and molecule, i.e., $\frac{1}{R}$, with $\frac{1}{R}+V_{\pi^{+}\pi^{+}}$, $\frac{1}{R}+V_{K^{+}K^{+}}$ and $\frac{1}{R}+V_{D^{+}D^{+}}$, respectively. From the numerical results, we see that the strong interactions between the two mesons almost do not affect the binding energies of the hydrogenlike molecular ion and molecule. The reason is that the electron(s) in the hydrogenlike molecular ion and molecule systems do not participate in the strong interaction, and these systems are bound by the electromagnetic attractive force of the electron on the mesons. The bond lengths for these systems are $2.003$ and $1.414$, respectively, in which case the strong interaction is negligible. So it does not affect the systems. 

\section{Summary}\label{sec5}  
Exploration of material structure is extremely important in physics, which helps us understand the interaction of fundamental elements in the entire world.  

In the present work, we propose three types of hydrogenlike structures by replacing the proton(s) in hydrogen atom, molecular ion and molecule with meson(s) of $\pi^{+}$, $K^{+}$ or $D^{+}$. In order to study these systems, we solve the Schr\"{o}dinger equation for hydrogenlike atoms, and use the variational method for hydrogenlike molecular ion and molecule. Since in the hydrogenlike matter, the masses of the mesons are all much heavier than that of electron, just like the case of hydrogen atom, molecular ion and molecule, the structures and the binding energies are almost the same as those of hydrogen systems. For $\pi^{+}e^{-}$, $K^{+}e^{-}$ or $D^{+}e^{-}$ system, the binding energy is $E_{n}=-\frac{1}{2n^{2}}$, and the most probable radius is $r_{n}=n^{2}$. For the systems of $\pi^{+}\pi^{+}e^{-}$, $K^{+}K^{+}e^{-}$ and $D^{+}D^{+}e^{-}$, there exist bound states if the spatial wave function is symmetric, while no bound states appear if the spatial wave function is antisymmetric. The binding energy of these hydrogenlike molecular ion is $-0.587$, and the bond length is $2.003$. For $\pi^{+}\pi^{+}e^{-}e^{-}$, $K^{+}K^{+}e^{-}e^{-}$ and $D^{+}D^{+}e^{-}e^{-}$ systems, if the spin wave function of the electrons is antisymmetric, and the spatial wave function is symmetric, there exist bound states. In this case, the two electrons can get close to each other, and form electronic cloud with high density near the midpoint of the two mesons, which provides electromagnetic attraction on the two mesons. As a result, the systems combine together. The binding energy of these systems is $-1.139$, and the bond length is $1.414$. Note that if the spin wave function is symmetric while the spatial wave function is antisymmetric, there is no bound state.

Comparisons of hydrogenlike molecular ion and molecule with the systems governed by the strong interaction are made in the following. In the hydrogenlike molecular ion, the two mesons repel each other, while the electron is attracted to each meson. This configuration is analogous to the doubly heavy triquark system, where the two heavy quarks form a color sextet, and this diquark together with the light antiquark form a color triplet. In such a color structure, the two heavy quarks experience a repulsive interaction, whereas the light quark and each heavy quark are mutually attractive. The existence of hydrogenlike molecular ions may suggest that doubly heavy triquarks could also exist, even though they cannot be directly observed in high-energy experiments due to not being color singlets.

Similarly, for hidden-flavor tetraquarks, if both the heavy quark pair $Q\bar{Q}$ and the light quark pair are in a color octet, the forces between the two heavy quarks and between the two light quarks are repulsive, while the interactions between heavy and light quarks are attractive. This pattern of repulsions and attractions mirrors that of the hydrogenlike molecule. On the other hand, in doubly heavy tetraquarks where $QQ$ is in a color sextet and $\bar{q}\bar{q}$ is in an antisextet, the interaction pattern again resembles that of the hydrogen-like molecule. Therefore, our findings on hydrogen-like molecular systems may offer insights into the possible existence of both hidden heavy-flavor and doubly heavy tetraquarks.

The hydrogenlike matter predicted in the present work has not yet received much attention, since its experimental observation remains challenging. With the improvement of accuracy and the development of detection technology in high energy physical experiments, these types of matter may be observed in the future, where the possible processes \cite{ParticleDataGroup:2024cfk} include $K_L^0\to \pi^+e^-\nu_e$ for $\pi^+e^-$ atom, $D^0\to K^-e^+\nu_e$ for $K^-e^+$ atom for instance.

\section*{Acknowledgments} 
This project is supported by the National Natural Science Foundation of China (NSFC) under Grants No. 11965016, 12335001, 11705069 and 12247101, and the Fundamental Research Funds for the Central Universities (Grant No. lzujbky-2024-jdzx06), the Natural Science Foundation of Gansu Province (No. 22JR5RA389, No.25JRRA799), and the `111 Center' under Grant No. B20063. 

Jun-Feng Wang and Zi-Yue Cui are the first and co-first authors, they contribute to this work equally. Cheng-Qun Pang and Zhi-Feng Sun are both the corresponding authors.

\section*{Appendix: Breit approximation, Fourier transformation and the function of $Y(\Lambda,m,r)$}
After obtaining the scattering amplitude of FIG. \ref{Fig5}, we use the Breit approximation to get the effective potential in the momentum space
\begin{eqnarray}
    \mathcal{V}^{AB\to CD}(\textbf{\textit{q}})=\frac{\mathcal{M}^{AB\to CD}(\textbf{\textit{q}})}{\sqrt{\prod_{i}2m_{i}\prod_{f}2m_{f}}}.
\end{eqnarray}
$m_{i}$ and $m_{f}$ are the mass of the initial and the final states. $\mathcal{V}^{AB\to CD}(\textbf{\textit{q}})$ and $\mathcal{M}^{AB\to CD}(\textbf{\textit{q}})$ are the effective potential in momentum space and the scattering amplitude of $AB\to CD$ process. The effective potentials in the momentum and coordinate space are simply connected by Fourier transformation,
\begin{eqnarray}
    \mathcal{V}^{AB\to CD}(\textbf{\textit{r}})=\int\frac{d^{3}\textbf{\textit{q}}}{(2\pi)^{3}}e^{i\textbf{\textit{q}}\cdot\textbf{\textit{r}}}\mathcal{V}^{AB\to CD}(\textbf{\textit{q}})F^{2}(q^{2}).
\end{eqnarray}
It is worth noting that the form factor $F(q^{2})$ is introduced to suppress the contribution of high momentum. In this paper, we take the exponentially parameterized form factor, i.e.,
\begin{eqnarray}
   F(q^{2})=e^{q^{2}/\Lambda^{2}}=e^{(q_{0}^{2}-\Vec{q}^{2})/\Lambda^{2}},
\end{eqnarray}
where $\Lambda$ is the cutoff. When performing the Fourier transformation, we will meet the following integral
\begin{eqnarray}
   Y(\Lambda,m,r)=\int\frac{d^{3}q}{(2\pi)^{3}}e^{i\Vec{q}\cdot\Vec{r}}\frac{1}{\Vec{q}^{2}+m^{2}-i\epsilon}e^{2(q_{0}^{2}-q^{2})/\Lambda^{2}}.
\end{eqnarray}
In Ref. \cite{He:2024aej,Sun:2024wxz}, we has calculated the $Y(\Lambda,m,r)$ function, which can be simplified as
\begin{eqnarray}
Y(\Lambda,m,r)&=&-\frac{e^{2q_{0}^{2}/\Lambda^2}}{(2\pi)^2r}\frac{\partial}{\partial r} \Bigg\{e^{2m^{2}/\Lambda^2}\frac{\pi}{2m}\Bigg[e^{mr}\nonumber\\
&&+e^{-mr}-e^{mr}\text{erf}\left(\frac{\Lambda r}{2\sqrt{2}}+\frac{\sqrt{2}m}{\Lambda}\right)\nonumber\\
&&+e^{-mr}\text{erf}\left(\frac{\Lambda r}{2\sqrt{2}}-\frac{\sqrt{2}m}{\Lambda}\right)\Bigg]\Bigg\}.
\end{eqnarray}

\newpage
\bibliographystyle{apsrev4-1}
\bibliography{reference}

\end{document}